\documentclass[sigconf]{acmart}

\AtBeginDocument{%
  \providecommand\BibTeX{{%
    \normalfont B\kern-0.5em{\scshape i\kern-0.25em b}\kern-0.8em\TeX}}}

\copyrightyear{2020} 
\acmYear{2020} 
\setcopyright{acmlicensed}
\acmConference[IUI '20]{25th International Conference on Intelligent User Interfaces}{March 17--20, 2020}{Cagliari, Italy}
\acmBooktitle{25th International Conference on Intelligent User Interfaces (IUI '20), March 17--20, 2020, Cagliari, Italy}
\acmPrice{15.00}
\acmDOI{10.1145/3377325.3377536}
\acmISBN{978-1-4503-7118-6/20/03}

\usepackage{enumitem}
\usepackage{hyperref}
\usepackage{xcolor}

\usepackage{titlesec}
\usepackage[normalem]{ulem}

\titlespacing*{\subsection}
  {0pt}{.3\baselineskip}{0pt}
\titlespacing*{\section}
  {0pt}{.3\baselineskip}{0pt}
  
\begin{document}

\title{\textit{\app}: Visual Counterfactual Explanations for Machine Learning Models}



\author{Oscar Gomez}
\authornote{Both authors contributed equally to this research.}
\author{Steffen Holter}
\authornotemark[1]
\email{{oscar.gomez, steffen.holter}@nyu.edu}
\affiliation{%
  \institution{New York University Abu Dhabi}
}


\author{Jun Yuan}
\author{Enrico Bertini}
\email{{junyuan, enrico.bertini}@nyu.edu}
\affiliation{%
  \institution{NYU Tandon School of Engineering}
}


\renewcommand{\shortauthors}{Gomez and Holter, et al.}
\newcommand{\app}{ViCE}

\makeatletter
\newcommand{\removelatexerror}{\let\@latex@error\@gobbl}
\makeatother

\begin{abstract}
  The continued improvements in the predictive accuracy of machine learning models have allowed for their widespread practical application. Yet, many decisions made with seemingly accurate models still require verification by domain experts. In addition, end-users of a model also want to understand the reasons behind specific decisions. Thus, the need for interpretability is increasingly paramount. In this paper we present an interactive visual analytics tool, \textit{\app}, that generates counterfactual explanations to contextualize and evaluate model decisions. Each sample is assessed to identify the minimal set of changes needed to flip the model's output. These explanations aim to provide end-users with personalized actionable insights with which to understand, and possibly contest or improve, automated decisions. The results are effectively displayed in a visual interface where counterfactual explanations are highlighted and interactive methods are provided for users to explore the data and model. The functionality of the tool is demonstrated by its application to a home equity line of credit dataset.
\end{abstract}





\maketitle

\section{Introduction}
The accessibility of high performing machine learning models has resulted in their integration into various applications pertaining to complex and high-risk data. Even in industries such as financial services and health care where the assimilation of predictive models has been slower due to the associated risks, machine learning is now seeing rapid adoption. However, simple accuracy measures that are used to describe models often fail to describe deeper flaws such as hidden biases and false generalizations. In high-risk situations such as cancer diagnosis or fiscal lending such oversight cannot be accommodated and can result in detrimental consequences.

In this paper we present \textit{\app}\footnote{\url{https://github.com/5teffen/ViCE}}, a novel design for an explainable machine learning visual analytics tool and its evaluation using a case study. The tool is built to describe a machine learning model by breaking down individual predictions. This caters directly to our end-user, who we envision as being the client-facing person trying to better understand predictions made by the model. This could include doctors inferring why a patient is predicted as high risk for diabetes or admissions officers looking into why a particular candidate was rejected. Although \textit{\app} could also be useful for model developers, that is not the tool's  primary purpose. The analysis is driven by the introduction of counterfactual explanations. Our tool relies on a new algorithm for calculating counterfactuals that is not limited to binary variables and is intended for use with tabular numerical data. Furthermore, we have created the first visual interface that is able to display these explanations effectively and coherently. \textit{\app} is supplemented with functionality that contextualizes the targeted sample with regards to the rest of the dataset. The combination of these features guarantees that the resulting interface does not only clarify the model's decision but can also be used to pinpoint bias and undesired behaviour.

For the targeted end-user each explanation provides actionable suggestions that can help adjust the model's prediction. In other words, the tool establishes what changes are required to alter their current state. For example, it could be used by a loan-officer looking to get a previously rejected application approved.
\section{Background and Related Works}
The problem of machine learning model interpretability and explanation has been recognized by many researchers and practitioners. Previous works \cite{biran2017explanation,doshi2017towards,lipton2016mythos,poursabzi2018manipulating,molnar2019,guidotti2019survey,adadi2018peeking,carvalho2019machine} provide an overview of methods, opportunities and challenges in this area. \looseness=-1

To interpret a machine learning model, methods vary according to the category of models. Machine learning models are often categorized into two classes: \textit{white-box} and \textit{black-box}. White-box models are those intrinsically interpretable models, where the logic of making a decision is transparent and intelligible (e.g. \textit{decision trees}, \textit{linear regression}, etc.) \cite{molnar2019}; while black-box models tend to have complex structures and are hard to understand (e.g., \textit{neural networks}, \textit{ensemble models}, etc.). In this paper, we introduce an explanation algorithm that is model independent, that is, the method works with any model without having access to its internal logic. \looseness=-1


Generally, model explanations can be categorized as \textit{local} or \textit{global}. Local explanations try to explain how a decision is made for a specific instance, while global explanation methods refer to showing the overall logical structure of a model. Some approaches such as LIME \cite{ribeiro2016should}, and SHAP \cite{lundberg2017unified}, focus on generating a weight for each feature as its contribution to the final decision. Others provide explanations through a counterfactual, where the explanation consists of the minimal set of changes to the feature values that allows the prediction for the instance to change to a different outcome. For example, finding the smallest feature perturbation that would change the prediction of a loan application from rejected to approved. Wachter \textit{et al.} \cite{wachter2017counterfactual} provide a general framework for counterfactual generation using stochastic optimization, while Ustun \textit{et al.} \cite{ustun2019actionable} present an approach specific to linear classifiers. \looseness=-1

As for the presentation of model explanations, visualization has been increasingly used to support the understanding, debugging, verification, and refinement of machine learning models. \looseness=-1


As a black-box explanation tool, \textit{\app} does not rely on the internal logic of the model, but is designed to let users explore the relationships between inputs (instances) and outputs (predictions). Some existing visual analytic systems follow the black-box approach. For example, Manifold~\cite{zhang2018manifold} enables the comparison of data distribution at different levels of granularity; RuleMatrix \cite{ming2019rulematrix} visualizes extracted rules for a given model; iForest \cite{zhao2019iforest} and Ensemble Matrix \cite{talbot2009ensemblematrix} attempt to explain ensemble models. \looseness=-1

Similarly to our work, the What-If Tool (WIT) \cite{wexler2019if} tries to answer a what-if question. WIT shows how model predictions change after the inference of data, while our tool, visualizes how to inference from data in order to change a prediction into other classes. \looseness=-1

Likewise, Rivelo \cite{tamagnini2017interpreting} and the Workflow for Visual Diagnostics proposed by Krause \textit{et al.} \cite{krause2017workflow} also provide a solution to a counterfactual question of how to change data to achieve a target class for black-box models. However, their solution adapts an algorithm \cite{martens2013explaining} originally designed for text documents and works only on binary inputs. Our work extends this algorithm to situations pertaining to continuous numerical data. \looseness=-1


\section{\app}

The main goal of the proposed tool is to support understanding of individual predictions through counterfactual explanations and to provide an intuitive visual representation for them. More precisely, our objective is to show, for a given instance, what is the minimal set of changes that is required to change the prediction. In our case, we focus on numeric features, therefore the tool has to provide two pieces of information: (1) which features need to change and (2) the extent to which they have to change. \looseness=-1

\textit{\app} was designed through an iterative design process. We analyzed published work to compile a list of questions end-users may want to answer when using counterfactual explanations and designed several solutions. The final result is the tool we present in the paper. The following list summarizes the desired functionality that we deemed essential to support our goal. \looseness=-1


\begin{itemize}[leftmargin=*, labelindent=10pt,topsep=0pt,nosep]
  \item[\textbf{Q1}] \textbf{Data distribution} - How do the values of the instance compare to those across the rest of the dataset? \\
  \textit{Example: If a student has a GRE score of 320, how does it compare to the scores of their peers?}
  
  \item[\textbf{Q2}] \textbf{Relevant features} - Which features have the most considerable effect on the model's prediction? \\
  \textit{Example: Identifying what variables in a patient's blood work are significant contributors to a negative diagnosis.}
  
  \item[\textbf{Q3}] \textbf{Possible changes} - Are there changes that could alter the model's current prediction? \\
  \textit{Example: If an applicant was rejected for a loan, what changes in their profile would be required for the application to be accepted?}

    \item[\textbf{Q4}] \textbf{Actionable changes} - Is it possible to change only a subset of \textit{actionable} features to change the model's prediction?\\
    \textit{Example: If a graduate school applicant knows certain features cannot be changed such as  \textit{Gender} or \textit{Age}, is it possible to generate an alternative explanation without altering these features?}
    
\end{itemize}

In the following two sections we first describe the algorithm developed in detail and then describe the visual solution designed for communicating information about the counterfactual explanations. \looseness=-1
\subsection{Counterfactual Algorithm}{0}
The counterfactual algorithm aims to find the minimal set of changes needed to change the model's output. We implement a simple heuristic algorithm to find changes that are at the same time interpretable (minimal set of features) and feasible (minimal amount of change); characteristics that are crucial for user-friendly explanations \cite{miller2019explanation}.\looseness=-1

In order to extend the algorithm proposed in \cite{martens2013explaining}, the entire dataset is discretized by fitting a Gaussian on each of the features and splitting the values into \textit{n} bins such that the middle \textit{n-2} capture four standard deviations from the mean, and the extreme bins capture data points beyond this. The algorithm greedily moves feature values across the bins until the predicted class is changed, or until the pre-defined constraints (no more than $w$ features are changed in a single explanation and no feature value is moved across more than $l$ bins) are reached.\looseness=-1

The algorithm starts with the original feature values of the instance to explain, and it is given an arbitrary set of unlocked features which can be acted upon. In each iteration, it independently moves the value in each of the unlocked features to the bins above and below the current one and chooses the one eliciting the largest change in the model's output (in the direction of the opposite category). It then takes the maximum change across all the unlocked features and uses this as the input for the next iteration. This greedy procedure continues until the modified instance crosses the model's decision boundary or until the constraints can no longer be satisfied.\looseness=-1



\subsection{Visual Interface}
\begin{figure}[h]
  \centering
  \includegraphics[width=0.6\linewidth]{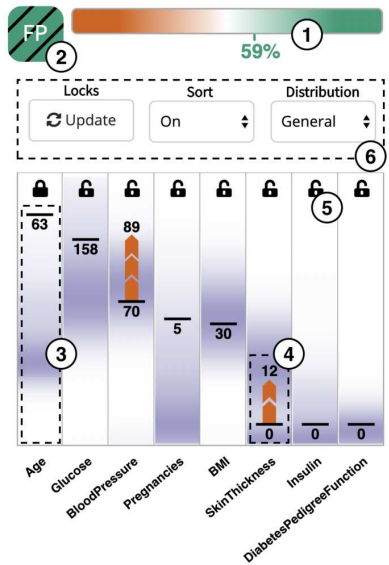}
  \caption{\textbf{Demonstrating a single local explanation using a diabetes dataset.} \textcircled{1} model’s predicted probability, \textcircled{2} classification correctness of the model’s prediction, \textcircled{3} frequency density distribution and the feature value for the given instance, \textcircled{4} counterfactual explanation, \textcircled{5} locking functionality, \textcircled{6} lock, sort, and distribution toggles.}
  \label{fig:diabetes}
\end{figure}

 In Fig. \ref{fig:diabetes}, we present the explanation for an instance in a diabetes dataset \cite{johannes1988using}. For demonstration purposes, a support vector machine is used. The individual explanation view shows a detailed summary regarding the model's decision for a single data point while also giving context to the values relative to the rest of the dataset. The percentage bar (Fig. \ref{fig:diabetes}\textcircled{1}) is used to indicate the exact prediction made by the model, thereby quantifying the strength of a prediction beyond the binary result. In our solution, any model prediction value greater than 50\% is classified as \textit{positive} and shown in green while all other decisions are classified as \textit{negative} and shown in red. The correctness of the prediction is also presented (Fig.~\ref{fig:diabetes}\textcircled{2}). It is important to note that knowledge of the ground truth is not a requirement as this information is unavailable in many real use-cases. However, when available, knowing whether the model's decision is correct helps categorize the sample point as either an example of the model's desired operation or of its potential shortcomings.\looseness=-1

The main part of the interface separates the data by features and displays their numerical values. These values are positioned relative to the distribution for that feature across the entire dataset (Fig. \ref{fig:diabetes}\textcircled{3}). Each attribute column is also supplemented with a density distribution visualisation. Based on the opacity of the purple background, the frequency of occurrence at that position can be analysed. For example, in this explanation the patient's age and glucose levels are clearly above the average. This information might suggest that these factors are contributing to the false positive prediction. By default, the tool displays the density distribution based on all the data points, however, the user has the option to map the densities based on points with \textit{positive} or \textit{negative} target values (\textit{Distribution} selection in Fig.~\ref{fig:diabetes}\textcircled{6}). In other words, it is possible to see how the sample under consideration compares with known \textit{positive} or \textit{negative} predictions. This effectively helps contextualize the values of the sample and highlight the features with singular values.\looseness=-1 

The local view will also display counterfactual explanations if the conditions set by the algorithm are fulfilled (Fig.~\ref{fig:diabetes}\textcircled{4}). Arrow shaped polygons are used to exhibit a single increment in the bins used to discretize the tabular data. The current value and the suggested new level are both shown numerically for clearer reading and detail. The color of these symbols is based on the binary decision made by the model. For a \textit{positive} prediction red arrows are used to show what changes would result in the decision becoming \textit{negative}, while green arrows are used for \textit{negative} instances as indicators for a \textit{positive} change. In this example, if two features had greater values then the patient would no longer be considered at risk for diabetes. Thus, according to the model, in order for this patient to become healthy they would need to slightly increase their blood pressure and skin thickness levels. This clearly exemplifies how the end-user benefits from having information that extends beyond a binary classifier. \looseness=-1

\begin{figure*}[h]
  \centering
  \vspace{-1em}
  \includegraphics[width=0.7\linewidth]{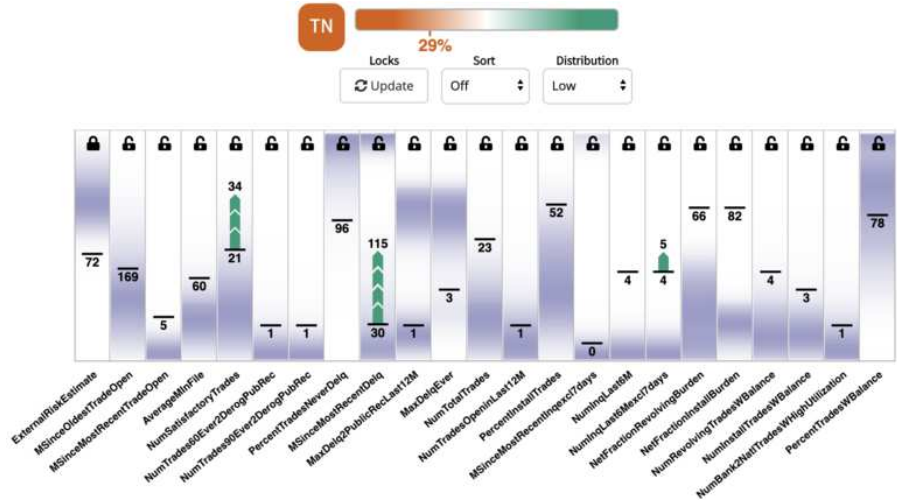}
  \caption{Use case for a sample from the HELOC dataset}
  \vspace{-1.5em}
  \label{fig:fico}
\end{figure*}


To guarantee versatility, a locking function is available to remove certain attributes from consideration (Fig.~\ref{fig:diabetes}\textcircled{5}). This can be useful if a user has certain features they deem unable to change or modify. In this case the age feature can be treated as unfeasible to change and can be locked using the icon. In most cases, the counterfactuals are elicited in examples with prediction percentages nearer to the cutoff threshold of 50\%. This is due to the fact that samples in which the model predicts a very high or low percentage usually cannot be flipped by implementing a few changes and would require larger modifications than those allowed by the algorithm.\looseness=-1

The tool also has a sorting option (\textit{Sort} in Fig.~\ref{fig:diabetes}\textcircled{6}). Toggling the sort button orders the features based on their standardized values. In this way users can quickly identify singular feature values that are considerably above or below the average for a feature. The sort functionality can be very effective in comparing monotonic features and highlighting key attributes.\looseness=-1

Notice that the lack of a counterfactual explanation does not mean that no information can be derived from the visualization. Comparing the data values to the density distributions depicted by the shaded area can help identify anomalies and derive hypotheses on why and how the model produces a given prediction. It is also worth noticing that the visualization interface can accommodate any other counterfactual generation methods \cite{ustun2019actionable, wachter2017counterfactual, laugel2017inverse, dice}.\looseness=-1


\subsection{Implementation}
\textit{\app} is built as a Flask web application with the back-end running on Python. The visualisations are created using D3 and JavaScript. With versatility in mind the tool accepts any binary classification dataset in a CSV format. A default SVM model is trained with scikit-learn, however, the program also accommodates custom input models as long as probability prediction methods are provided. The data is processed in real time to accommodate customized end-user inputs. For our implementation we split feature values across $n = 10$ bins and set $w = l = 5$ for the algorithm constraints.\looseness=-1

\section{Case Study}
To demonstrate how \textit{\app} can help with ML explanation, we showcase its use with the Home Equity Line of Credit (HELOC) dataset. The design goals set out in Section 3 are used to evaluate the performance of the tool and are referenced directly in the use case. 

The HELOC dataset was released as part of the FICO xML challenge~\cite{fico}. It is comprised of applications made by real homeowners in attempts to get a credit line from the bank. The target is to predict the binary variable \textit{Risk Performance} where \textit{bad} indicates that a consumer was at least 90 days past due once and \textit{good} that they never were. Some initial testing revealed that the \textit{External Risk Estimate} feature had a very strong correlation with the target class. Since this feature is not directly actionable it is initialized as locked. However, the user retains the ability to unlock it if desired. \looseness=-1



To simulate the end-user experience an arbitrary client was picked. The chosen instance seen in Fig.~\ref{fig:fico} shows a negative model prediction with 29\% probability and the TN label on the upper left indicates that the model prediction matches the ground truth. Setting the data point into context using the density distributions reveals that there is considerable variation from the dataset averages in a number of the features \textbf{(Q1)}. Toggling the sort functionality helps identify the most singular features to be \textit{Net Fraction Revolving Burden}, \textit{Percentage Trades Never Delinquent} and \textit{Months Since Most Recent Delinquency}. The \textit{External Risk Estimate} score is also considerably lower than the average. Since this sample elicits a number of uncommon values there is a certain degree of subjectivity involved with identifying the features that should be of particular interest. Redrawing the distribution for \textit{negative} samples shows the frequency density distribution for other known \textit{negative} points. Using this view it is possible to confirm that the features identified above are significantly different, even in the context of other poorly performing samples \textbf{(Q2)}. Furthermore, changing between the general, positive, and negative density frequency distribution views gives an indication of the monotonicity of the features. \looseness=-1

To understand what changes would be required by the user to receive a positive prediction we can examine the counterfactuals. This sample cannot be considere an edge case, however, since the percentage prediction is not too low at 29\% and there still exist combinations of changes that would result in the model flipping the decision \textbf{(Q3)}. Yet, as expected, these changes are significant. The tool suggests that a sizable increase in both the \textit{Number of Satisfying Trades} and \textit{Months Since Most Recent Delinquency} and a small rise in the \textit{Number of Inquiries in the Last 6 Months excluding last 7 days} would be sufficient. Intuitively, all of these changes are manageable, but if the user was in a rush to get their credit line approved the time based features might not be feasible \textbf{(Q4)}. Locking these attributes and reloading the explanation generates a new explanation with changes in \textit{Average Months in File}. Since this is also time dependent it was subsequently locked as well. With these limitations imposed by the user, the algorithm is no longer able to identify a way in which the decision can be changed within the pre-defined constraints. Therefore, it is apparent that the model weights features with time variables highly in its decision making for this instance. For further exploration, unlocking the \textit{External Risk Estimate} variable instantly demonstrates the strength of its correlation with the model decision. The explanation now suggests large changes in \textit{External Risk Estimate} as the optimal way of flipping the decision. \looseness=-1
\section{Limitations}
This work is the first step in our goal to provide full end-user oriented model explanations. The tool currently has certain limitations. For example, the algorithm cannot effectively handle categorical features. Possible solutions might involve presetting a search path or performing a brute force analysis of features that are known to be categorical. In addition, the tool does not extend to multi-class classification or other contexts such as image classification. The visualization itself can realistically display a maximum of around 30 features. However, larger datasets can be accommodated by utilizing the sorting feature and only displaying the top \textit{k} features. \looseness=-1
\section{Conclusions and Future Work}
In this paper, we presented \textit{\app} – a novel way for the end-user to gain insight into model predictions through counterfactual explanations. For each sample the minimal set of changes needed to alter the decision was shown. Interacting with the interface allows customizing the explanation according to the user's requirements. A use case was chronicled by applying the tool on a loan dataset. To the best of our knowledge this tool is the first in visualising counterfactuals for non-binary data. While already providing increased model interpretability, the modular black-box based nature of the tool allows for a seamless integration of improvements such as including different methods to generate counterfactuals, or providing users with a set of alternatives to the displayed counterfactual explanation. \looseness=-1

Future work will aim to introduce increased interactivity for the UI. This would include adding an option to view the impact of custom changes inputted by the user. To improve the visualization, additional explanation methods can be integrated. For example, customizing the sorting functionality to order the features according to their local importance magnitudes could provide a way to corroborate the insights gained from the counterfactuals. Finally, extending the tool to a global scale through the aggregation of instance explanations could further increase its usefulness for model developers. \looseness=-1
\section*{Acknowledgements}
This work was partially supported by the DARPA D3M program. Any opinions, findings, and conclusions or recommendations expressed in this material are those of the authors and do not necessarily reflect the views of DARPA. We thank the nexquare team, especially Haseeb Ahmed and Sarah Ameen, for providing valuable feedback and sharing their unique perspective, having led multiple data science engagements in the education domain. We also thank the anonymous reviewers for their valuable feedback. \looseness=-1

\bibliographystyle{ACM-Reference-Format}
\bibliography{main}
\end{document}